\documentclass[sigplan,10pt]{acmart}
\pdfoutput=1

\renewcommand\footnotetextcopyrightpermission[1]{} 
\setcopyright{none}

\usepackage{textcomp}
\usepackage{fancyhdr}
\usepackage{amsmath}
\usepackage{algorithm}
\usepackage{algorithmic}
\usepackage{float}
\usepackage{subfigure}
\usepackage{multirow}
\usepackage{rotating}
\usepackage[inline]{enumitem}
\usepackage{calc}
\usepackage{flushend}
\usepackage{url}
\usepackage[]{hyperref}

\usepackage{graphicx}

\floatstyle{plain}
\newfloat{myfloat}{thp}

\usepackage{listings}
\usepackage{courier}
\usepackage{color}
\usepackage{booktabs}
\usepackage{colortbl}
\usepackage{tikz}
\usetikzlibrary{shapes,arrows,decorations.markings,fit,backgrounds}
\usepackage{flowchart}

\tikzstyle{decision} = [diamond, draw, fill=red!10,
    text width=9.5em, aspect=4, text badly centered, node distance=1cm, inner sep=-2pt]
\tikzstyle{block} = [rectangle, draw, fill=blue!20,
    text width=15em, text centered, rounded corners, minimum height=1em]
\tikzstyle{line} = [draw, -latex']

\definecolor{sh_comment}{rgb}{0.00, 0.50, 0.00 } 
\definecolor{sh_keyword}{rgb}{0.00, 0.00, 0.60 }  
\definecolor{sh_string}{rgb}{0.40, 0.20, 0.40 } 

\usepackage{colortbl}

\definecolor{evencolor}{gray}{0.8}
\definecolor{oddcolor}{gray}{0.99}
\definecolor{LightGray}{RGB}{228,228,228}

\newcommand{\refSec}[1]{Section~\ref{#1}}
\newcommand{\refSecs}[2]{Sections~\ref{#1}~and~\ref{#2}}
\newcommand{\refFig}[1]{Figure~\ref{#1}}

\newcommand{\refFigss}[3]{Figures~\ref{#1},~\ref{#2} and~\ref{#3}}

\newcommand{\refTab}[1]{Table~\ref{#1}}

\newcommand{\refLst}[1]{Listing~\ref{#1}}

\newcommand{\code}[1]{{\ttfamily #1}}

\newcommand{\nameShort}{ZOFI}

\begin{document}
\title[{\nameShort}]{{\nameShort}: Zero-Overhead Fault Injection Tool for Fast Transient Fault Coverage Analysis}
\author{Vasileios Porpodas}
\affiliation{
  \institution{Intel Corporation}
}
\email{vasileios.porpodas@intel.com}
\begin{abstract}
  The experimental evaluation of fault-tolerance studies relies on tools that inject errors while programs are running, and then monitor the execution and the output for faulty execution.
  In particular, the established methodology in software-based transient-fault reliability studies, involves running each workload hundreds or thousands of times, injecting a random bit-flip in the process.
  The majority of such studies rely on custom-built fault-injection tools that are based on either a modified processor simulator, or a code instrumentation framework.
  Such tools are non-trivial to develop, and are usually orders of magnitude slower than native execution.

  In this paper we present {\nameShort}, a novel timing-based fault-injection tool that is aimed at being used in fault-coverage studies for transient faults.
  {\nameShort} is a zero-overhead tool, meaning that the analyzed workload runs at native speed.
  This is orders-of-magnitude faster compared to common approaches that are designed around simulators or code instrumentation tools.
{\nameShort} is free software and is available at \url{https://github.com/vporpo/zofi}.
\end{abstract}

\maketitle

\section{Introduction and Related Work}
\label{sec:introduction}

Transient faults, also known as soft errors, are faults that occur once and do not persist~\cite{sorin2009fault}.
They are a major cause of reliability issues, as it has been shown in several studies~\cite{constantinescu2003trends, michalak2005predicting, shivakumar2002modeling, srinivasan2004impact}.
These faults can be attributed to a range of factors, including alpha particles striking the silicon circuits, fluctuations in the power supply and others. 
Such events can cause bit-flips in digital circuits, which can corrupt the state of the logic.
A transient fault in a processor can change the outcome of instructions, possibly leading to wrong execution, which can potentially cause a system crash.
Transient faults are fairly frequent in large data centers~\cite{michalak2005predicting}, due to the large number of systems, and therefore such systems have to be designed with transient faults in mind.

Software-based error detection techniques, e.g.,~\cite{chang2006automatic,feng2010shoestring,oh2002error,Reis2005,Wang2007,Zhang2010,drift,mitropoulou2013casted,mitropoulou2016comet,Shye2007using,ghosh2012runtime}
, aim at providing reliability at the software level.
Such techniques work by replicating the code at either the instruction, the thread or the process level, in an attempt to detect transient faults.
The evaluation of their effectiveness is done with the help of fault-injection tools which apply random register bit-flips and check the result for either: (i)~successful detection by the technique, (ii)~data corruption, (iii)~infinite execution, (iv)~exception, or (v)~correct execution, also known as masked errors.

\subsection{Types of Fault-injection Tools}
\label{sec:introduction:types}
Fault-coverage evaluation for transient faults implements the single-event upset model.
It requires the use of a tool that runs the target workload, pauses it, injects a bit-flip, resumes its execution and then tracks its execution to check for a faulty outcome.
In the error detection literature this is commonly done in various ways:
\begin{enumerate}

\item {\bf Source Code Editing:} Delete, or modify instructions in the source code or in the compiler intermediate representation.
This is a straight-forward approach, but it does not accurately model transient faults.
An example of a study that uses this is~\cite{Oh2002}.

\item {\bf Simulator/Emulator:} Modify a processor simulator such that it can inject a fault at a given execution cycle. 
Depending on the accuracy of the simulation, this can vary from extremely slow (e.g., at the RTL-level), to quite slow (tens/hundreds of MIPS) for the fastest emulators.
This is a very popular technique and has been used in several studies, including \cite{Feng2011,khudia2013low,mitropoulou2013casted,drift,khudia2014harnessing,didehban2016nzdc,sartor2017exploiting,so2018expert,kaliorakis2015differential,chang2018hamartia,fsefi}.

\item {\bf Binary, Source, or Compiler-IR Instrumentation:} Build a custom binary-instrumentation tool using a framework like Pin\cite{luk2005pin}, and implement fault-injection with it, or use a tool based on this approach, like LLFI~\cite{llfi}, PINFI~\cite{pinfi} and SASSIFI~\cite{sassifi}.
The instrumentation overhead is still quite significant, leading to execution much shower than the non-instrumented binary.
A few examples of studies using this approach are \cite{Reis2005,Chang2006,Wang2007,Shye2007using,rink2017extending,rink2017flexmedic,davidson2018error,previlon2019pcfi,kestor2018understanding,li2018modeling,georgakoudis2017refine,georgakoudis2019safire}.

\item {\bf Timing-based Tool:} Run the unmodified binary on native hardware, just like with a debugger, pause the execution, inject the fault and resume the execution at native speed.
This is by far the fastest of the three, and as we explain in the text, it is also equally accurate for statistical fault coverage analysis as long as micro-architectural accuracy is not required.
We are aware of a relatively small number of studies that use this approach~\cite{ghosh2012runtime,mitropoulou2016comet,vankeirsbilck2017random,thati2018instruction,esift,chen2018ladr}.
In these studies a custom in-house tool was developed, usually designed around a debugger, like GDB~\cite{gdb}.
\end{enumerate}

In this paper we present {\nameShort}, a complete fault-coverage analysis tool, based on the time-based design, that executes the unmodified binary on native hardware with no extra overheads whatsoever.
It executes the test binary at full native speed, then pauses it a given time for the fault injection to occur.
Once the binary is stopped, the {\nameShort} tool can modify the state of the target workload by injecting a fault, implemented as a register bit-flip, and then resumes its execution.
The workload will either execute to completion, or will get interrupted by a signal.
At that point {\nameShort} compares its output against the original run, and determines the type of the execution outcome.
In case of an infinite execution, {\nameShort} will force-stop the workload.
After repeating this process for a large number of such test runs and collecting statistics data, the final results are summarized and reported to the user.

The {\nameShort} tool is free software, distributed under the GPL version 2.0 license.
The project is hosted at \url{https://github.com/vporpo/zofi} and is implemented in C++.
The tool is designed for modern x86\_64 Linux-based systems.

\section{Background and Motivation}
\label{sec:background}
This section presents an overview of the established fault coverage evaluation methodology (\refSec{sec:background:overview}), and motivates the need for timing-based injection tools, like {\nameShort}, in \refSecs{sec:background:slow}{sec:background:benefits}.

\subsection{Overview of the Fault-Coverage Evaluation Methodology for Transient Faults}
\label{sec:background:overview}

\begin{figure*}[h]
\centering
\includegraphics[width=4.9in]{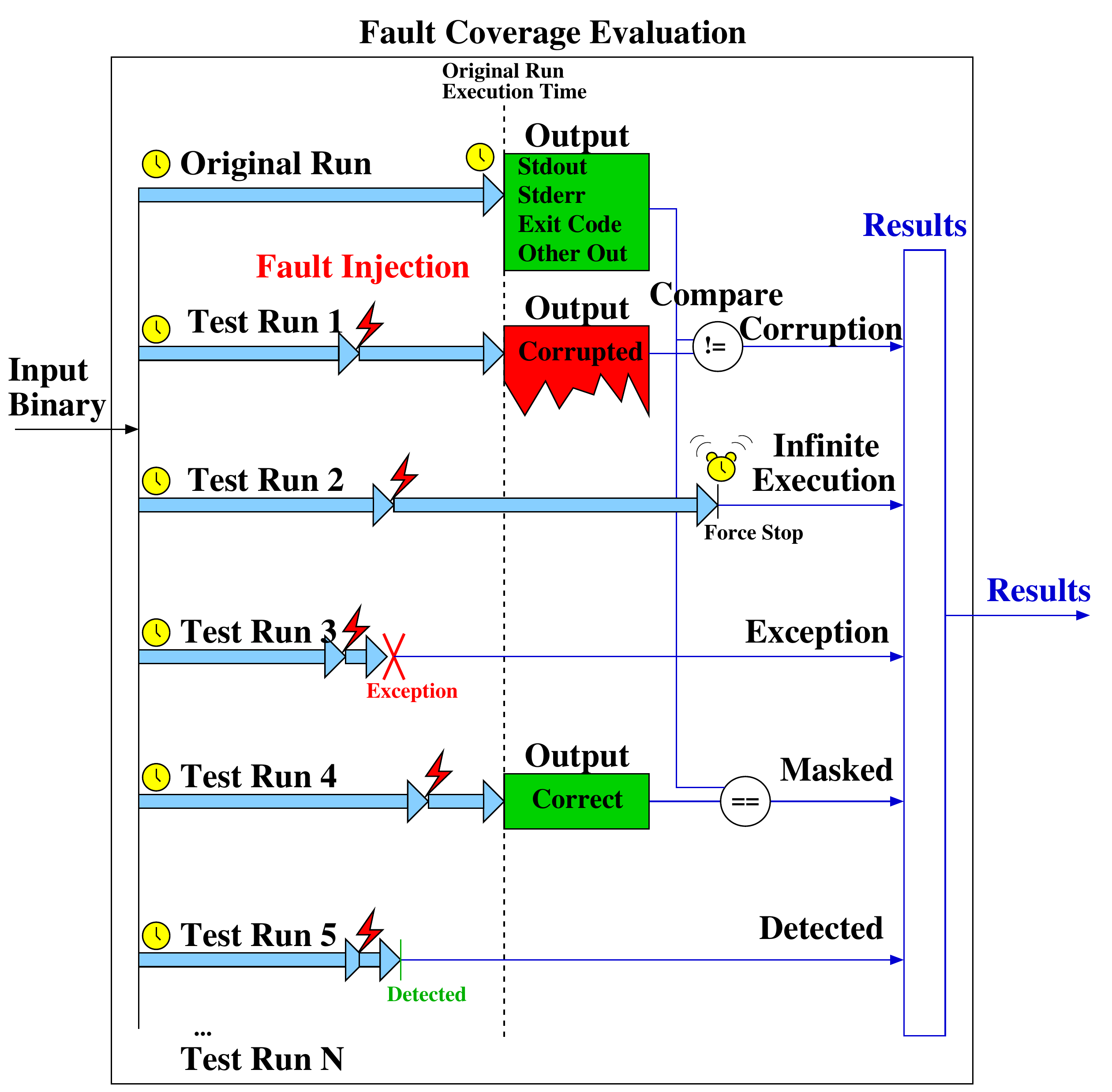}
\caption{A typical fault coverage evaluation process.}
\label{fig:fault_coverage_evaluation}
\end{figure*}

An overview of a typical fault coverage evaluation of a binary is shown in \refFig{fig:fault_coverage_evaluation}.
The input to the process is the binary that we are analyzing, along with its inputs, which are not shown in the figure.

The first step is to run the binary, measure its execution time (or cycles) and collect all its outputs.
This includes \code{stdout}, \code{stderr}, exit code and all output files.
We refer to this initial run as the ``Original Run'', and it is shown as the top-most run in \refFig{fig:fault_coverage_evaluation}.
The execution time (cycles) of the original run is used:
\begin{enumerate}
\item as an upper bound to the fault injection time (cycle), and 
\item as an estimate of how long the analyzed workload should take to complete, such that we can detect and force-stop infinite executions.
\end{enumerate}
The original output is used to check the output of the test runs, as we will explain shortly.

The rest of the process involves a large number of ``Test Runs'' which is where fault injection takes place.
Each of these runs executes the same binary with the same inputs, but their state gets altered by a fault-injection event (depicted as a red lightning in the figure), at some point during their execution.
The injection of the fault takes place at a random point in time (cycle) between zero and the execution time (cycle) of the original binary.
The type of fault injected depends on our configuration, but a common fault type is a bit-flip into registers to simulate transient faults in the computational logic.
After fault injection the binary resumes its execution and is let free to execute to completion.

Now let's take a look at the fault injection runs.
In ``Test Run~1'' of \refFig{fig:fault_coverage_evaluation}, the fault leads to output (the red deformed block) that differs form that of the original run.
This is caught by the comparator component, which reports a ``Corruption''.

Another example of a test run is ``Test Run~2'', where the state of the execution gets damaged by the fault in such a way that leads to infinite execution.
This can be caused, among others, by faults that lead to a faulty evaluation of the condition bit of a loop latch.
If the execution of the binary takes far longer than the original execution time, an alarm goes off that forces the workload to stop.
This is reported as ``Infinite Execution''.

A third example is ``Test Run~3'', where the fault leads to an illegal operation that triggers an exception.
Common examples are: memory instructions that attempt to access memory outside their allocated memory, floating point exceptions and others.
This is reported as an ``Exception''.

``Test Run~4'' shows an example where the injected fault does not cause any failure whatsoever.
The test run executes to completion, its output is identical to the original one, and it did not trigger any exception.
This can happen when the bit affected by the fault remains unused by subsequent computation.
For example, it may be overwritten or discarded by the rest of the execution.
This type of run reports is reported as ``Masked'' (also known as ``Benign'').

Finally, in error detection studies, the system itself has a way of detecting errors and reporting them to the fault-injection framework.
This is shown in ``Test Run~5''.
The error detection system can notify the injection tool that it has detected an error in several ways:
For example, it can return a specific exit code, or it can print specific output to \code{stderr}.
This is reported as ``Detected''.

\subsection{Fault Coverage Analysis is Time Consuming}
\label{sec:background:slow}
Fault coverage is a Monte-Carlo statistical analysis that relies on a large number of experiments.
In each experiment we run the workload and we inject a random fault into it at a random time.
Thus, the time it takes to run all these experiments is comparable to running all these workloads to completion.

Achieving good accuracy, requires us to run the workload hundreds or thousands of times.
Therefore, if our fault-coverage analysis tool introduces a large performance overhead, for example by using a simulator for running the workload, then the analysis becomes impractical even for workloads that would run relatively fast on native hardware.
The {\nameShort} timing-based injection tool is designed to make such analysis practical, even for long-running workloads, by removing the overhead introduced by the fault-injection tool.

\subsection{Timing-based Tool: No Need for Cycle Accuracy in Fault-Coverage Analysis}
\label{sec:background:benefits}

\begin{table*}[h]
\begin{center}
  {
    \caption{High-level comparison of fault injection tool designs.}
    \begin{tabular}[t]{l|l|l|l|l}
      \toprule Type   & Speed & Granularity & Injection Accuracy & Access to u-arch\\
      \midrule
\rowcolor{evencolor}        Source Code Editing        & Native & Instruction & Fixed (Not a Transient Fault) & No \\
\rowcolor{oddcolor}        Simulator (cycle accurate) & Low & u-Op         & Cycle (High) & Yes \\
\rowcolor{evencolor}        Emulator (functional)      & Med & Instruction   & Dynamic Instr. Count (Med) & No \\
\rowcolor{oddcolor}        Binary Instrumentation  & Med & Instruction   & Dynamic Instr. Count (Med) & No  \\
\rowcolor{evencolor}Timing-based ({\nameShort}) & Native & Instruction & Time (Low)        & No  \\
      \midrule
    \end{tabular}

    \label{tab:designs}
  }
\end{center}
\end{table*}

As explained in \refSec{sec:introduction:types}, there are several types of fault injection tool designs, which are summarized in \refTab{tab:designs}.
Each of them operates at different speeds, has different fault-injection granularity and accuracy, and may or may not have access to micro-architectural components.
In this work, we argue that for most practical use-cases the timing-based design is the best option, because it provides the maximum possible test speed without sacrificing the accuracy of our statistical analysis.

Since fault coverage is a statistical analysis that does not require cycle accuracy, there is no real need to sacrifice execution speed for cycle accuracy.
Therefore, we believe that the proposed timing-based design of {\nameShort} should be the choice of preference for most studies, similarly to sampling-based profiling tools (e.g.,~\cite{perf,vtune}) which are adequate for many profiling scenarios.

Simulator-based tools give us access to the micro-architectural state, which is not accessible by the other techniques.
However, this comes at the cost of a huge overhead in execution speed, limiting its practical uses to small execution kernels or traces.
We believe that for any fault-coverage study that does not need access to micro-architectural state, it is more preferable to use any of the alternative techniques.

Therefore, once the cycle accuracy is not required, one could use any of the other designs interchangeably, with an equivalent outcome.
None of the emulators, the instrumentation-based, or the timing-based tools have access to the micro-architectural components, as they are all limited to the state exposed by the ISA.
Moreover, while emulators and instrumentation-based tools do have an instruction-based injection accuracy (as they can stop at precisely a specific number of dynamic instructions), we argue that this is not really needed for statistical analysis, as the faults get injected at random cycles anyway.
Therefore, we can safely use a random time instead of a random dynamic instruction, and this should give us equivalent results.
As a result, timing-based fault injection tools, like {\nameShort}, are as effective as emulators and instrumentation-based tools, in their fault injection analysis, but with the additional benefit of being orders of magnitude faster.

\section{{\nameShort} Design}
\label{sec:proposed}

This section describes {\nameShort} in more detail. 
It provides a description of the design and implementation, and lists a set of important features.

\subsection{Overview}
\label{sec:proposed:overview}
Just like any other fault-coverage analysis tool (as shown in \refFig{fig:fault_coverage_evaluation}), the high-level steps followed by the {\nameShort} tool are the following:
\begin{enumerate}
\item Run the workload with no fault injection to measure its execution time and collect its original output. We refer to this as the ``Original Run''.
\item Fork and spawn a new process for the test run of the workload. This is where fault injection will take place.
\item In the meantime, pick a random time point between zero and the execution time of the original run. This will be the time point when the fault injection will take place. Sleep for this amount of time.
\item Wake up and stop the test process. Analyze the currently executed instruction and randomly pick a register. The type of register to be picked is controlled by the user options provided.
\item Flip a random bit of that register.
\item Let the process continue to completion.
\item Collect the output and compare it against the original run.
\end{enumerate}

\subsection{Design}
\label{sec:proposed:design}

{\nameShort} is designed around the \code{ptrace} Linux system call, that provides the means by which one process (the {\nameShort} tool) can observe and control the execution of another process (the workload).
This is commonly used by debuggers for observing, controlling and updating the state of the debugged binary.
The \code{ptrace} calls give you full access to the register state, the memory and the code.
It is therefore a perfect fit for our fault injection tool.

\begin{figure}[h]
\centering
\includegraphics[width=2.9in]{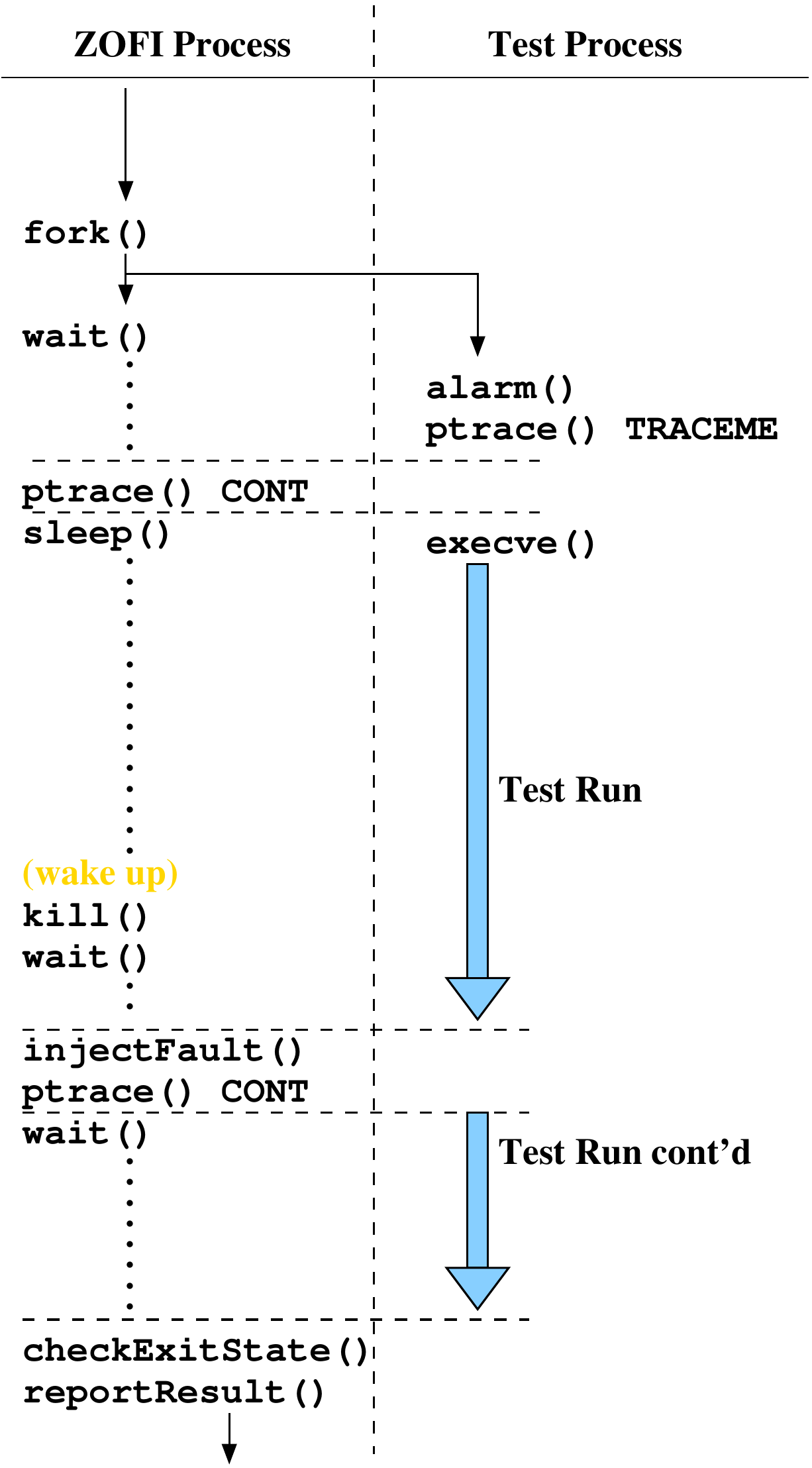}
\caption{The design of the fault-injection run.}
\label{fig:design}
\end{figure}

Fault injection works as follows.
The execution of the workload is interrupted by a signal emitted by the main process of the tool, and its register state is imported into the tool.
So at this point, {\nameShort} has access to the instruction pointer, but does not know what type of instruction this is, in order to analyze it and access the registers it reads/writes.
The analysis of the instruction is done using the capstone~\cite{capstone} library.
This provides the list of the accessed registers, the type of each access (read/write), whether this is an explicitly or implicitly accessed register and other information, giving the user a good level of control on the type of registers that are targeted by faults.

An overview of the design of the fault-injection part of {\nameShort} is shown in \refFig{fig:design}.
The tool process (tracer) forks a new child process which becomes the tracee.
Initially the child process sets up an alarm so that it receives a signal after some time if it gets stuck in an infinite execution loop.
After an initial synchronization, the tool process sleeps for a random time until it is time for the fault injection.
Meanwhile the child process switches to the target binary image with a call to \code{execve()}.
Once the tracer process wakes up, it stops the tracee with a \code{kill()}, and after a synchronization point it performs the fault injection.
Then it lets the tracee continue, and waits for it to either run to completion, or stop with an exception, or never complete due to infinite execution.
After the final synchronization point, the tool's main process compares the exit state of the test run with that of the original run and reports the result to the results-collection part of the tool.

\subsection{Modeling of a Transient Fault}
\label{sec:proposed:injection}
\begin{figure}[t]
\centering
\includegraphics[width=3.4in]{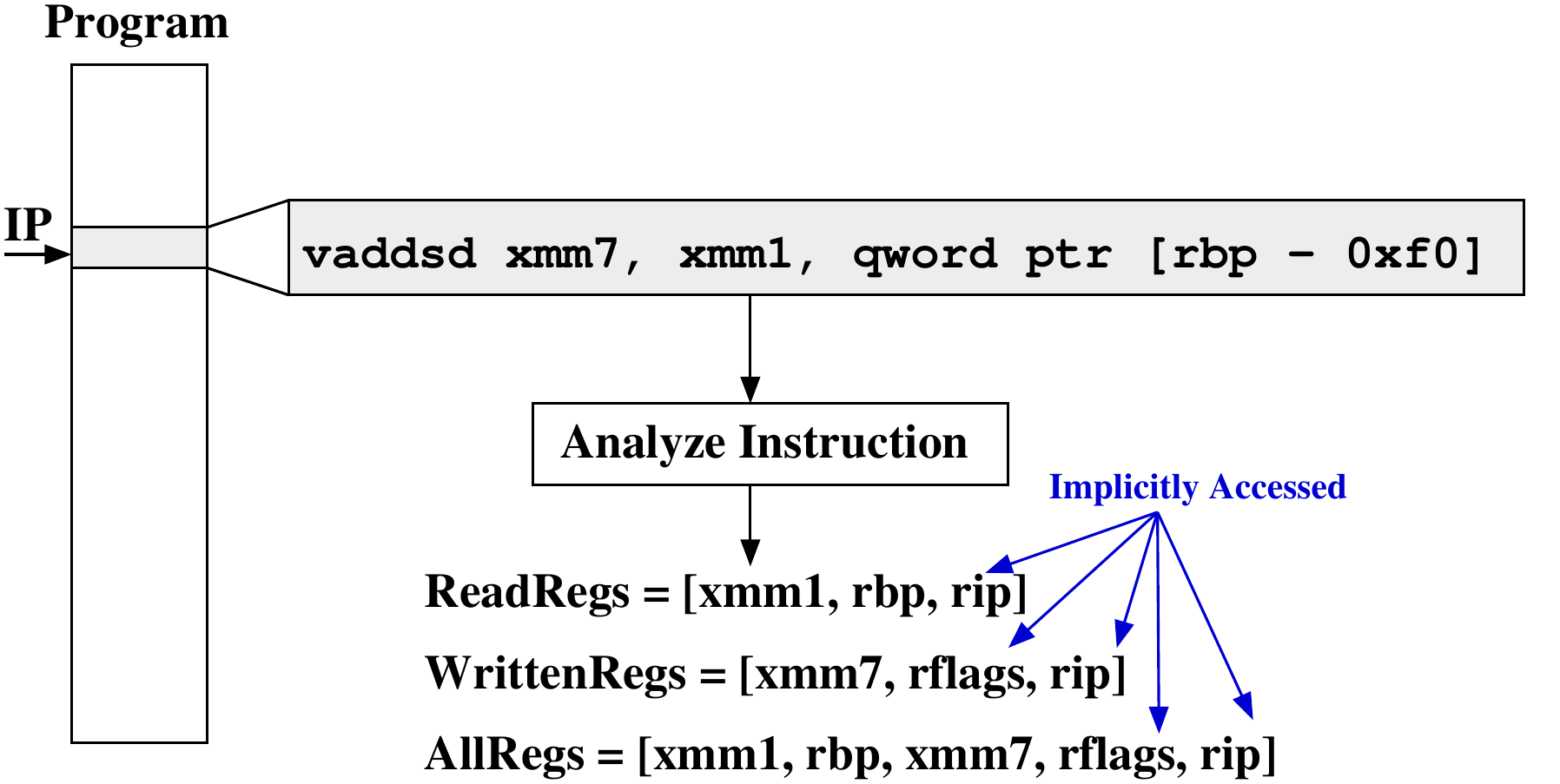}
\caption{Analyzing the instruction at the current Instruction Pointer (IP), to collect the accessed registers for each register type.}
\label{fig:injection}
\end{figure}

We model the transient fault in a similar way as in most fault injection studies: with a random bit-flip in a random register from those accessed by the current instruction.
Currently, {\nameShort} implements a register bit-flip injection, since this is the most widely used model in the studies, but this could be easily extended to include bit-flips in memory.
The reasoning is that ECC memory is commonly used in server-grade hardware, and therefore such systems can automatically recover from transient faults in memory cells.

The random register bit-flip is implemented as follows.
After the child process gets stopped by the parent, the instruction pointer points to the instruction that is about to be executed.
{\nameShort} analyzes the instruction using the capstone library~\cite{capstone}, and collects the registers read and written either explicitly or implicitly (for example the instruction pointer, or the condition flags in x86).
An example of this is shown in \refFig{fig:injection}, where an instruction is analyzed and the registers accessed are collected into three pools: one for the registers being read, one for those being written, and one for all registers being accessed.
Depending on the user's input, {\nameShort} may filter out some of these registers, for example the user may wish to skip the implicitly accessed registers, or collect the instruction pointer only when the instruction is a control-flow one, or even restrict the errors to registers read or written.

\begin{figure}[h]
\centering
\includegraphics[width=3.4in]{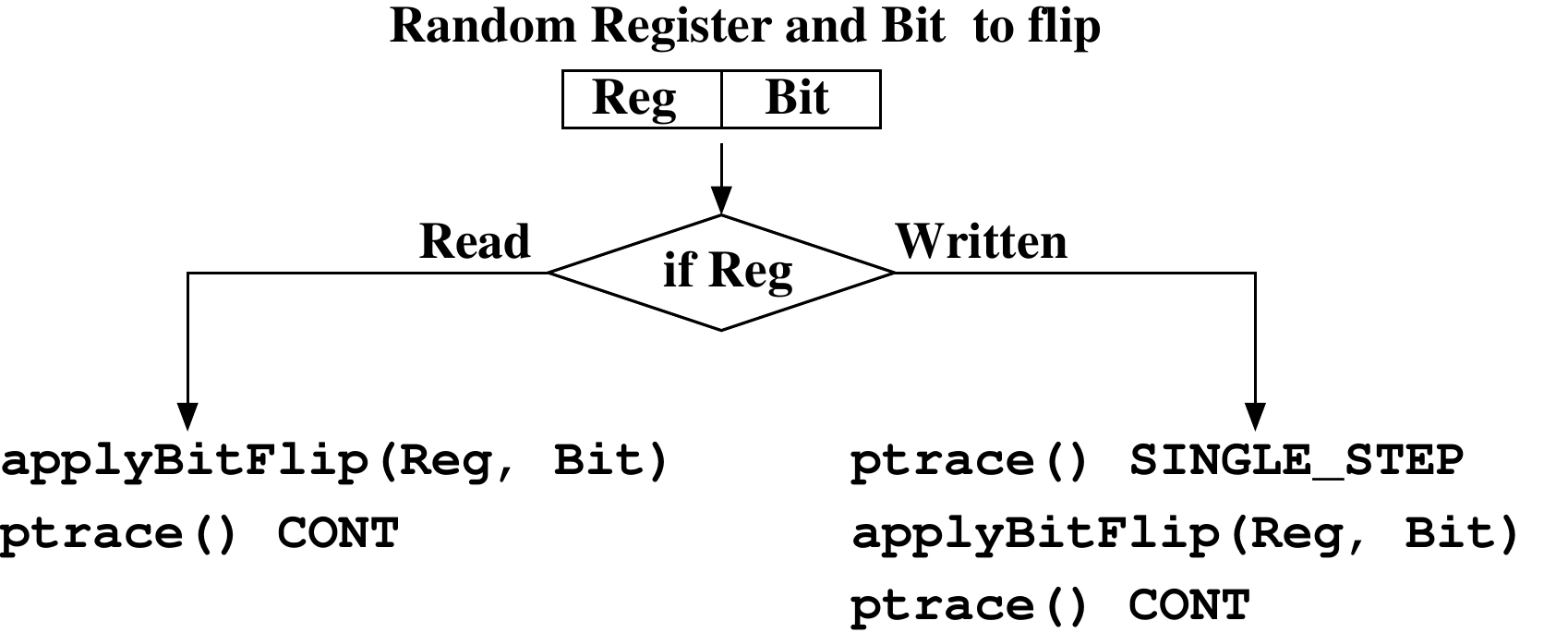}
\caption{The injection process is different when the register is read from when it is written.}
\label{fig:injection_r_w}
\end{figure}

The next step is to randomly select one of the registers from the corresponding pool.
Once the register is selected, {\nameShort} randomly selects a bit where the bit-flip should take place.
If we are injecting errors only into registers read by the instruction, we are now ready to apply the bit-flip and continue the execution, as the bit-flip will be processed by the current instruction and will flow through the execution logic.
This is shown on the left hand side of \refFig{fig:injection_r_w}.
If, however, we are injecting errors into the output (written) registers of the instruction, we cannot follow the same process, as a bit-flip performed while the child process is stopped, will be overwritten by the output of the current instruction once it gets executed.
Therefore, in this case, {\nameShort} will first step into the next instruction and will then apply the bit-flip.
This is shown on the right hand side of \refFig{fig:injection_r_w}.
Next, the execution of the tracee process continues.

\subsection{Features}
\label{sec:proposed:features}
In this section we present a list of {\nameShort}'s most important features:
\begin{itemize}
\item Execution of test runs at native speed, as there is no emulation or instrumentation involved.
\item Support for multiple concurrent test runs (jobs) using the \code{-j N} flag. In this way the test throughput increases almost linearly to the number of cores.
\item Built-in tracking of the execution state, output checking and collection of statistics.
\item The user can fully override the comparator function that checks the output. The user-defined comparator runs in a system shell.
\item Support for multi-threaded workloads (since v0.9.4).
\item Option to avoid injecting faults into the system's library code. 
This is particularly helpful for accurate measurements of workloads that are protected by fault-tolerance techniques linked against non-protected system libraries.
\item Fine control of where the faults will be injected.
\item Modular design and implementation in modern C++ and well documented code. Easy to modify and extend.
\end{itemize}

\subsection{Limitations}
\label{sec:proposed:limitations}
The main limitation of {\nameShort} is not due to its implementation, but rather due to its timing-based design.
If the target workload runs for a very small amount of time, then {\nameShort} won't be able to effectively inject faults, often attempting to inject the fault after the workload has finished executing.
This is a common problem to all timing-based tasks on real systems. 
For example, even measuring the absolute execution time of a workload is not reliable when the execution lasts for a very short time.
When we get close to the time accuracy of the system, any timing-based functionality becomes unreliable.
We observed that it is best if the workload runs for at least tens or hundreds of milliseconds.

Another limitation is the evaluation of workloads that misbehave when being stopped with a signal.
{\nameShort} relies on signals to pause the applications, therefore it cannot properly analyze such types of workloads.

\section{Results}
\label{sec:results}
This section provides a quantitative evaluation of {\nameShort} by presenting a set of results on: (i) its performance overhead compared to a native run (\refSec{sec:results:performance}), (ii) a fault-coverage analysis of the NAS NPB-2.3\cite{nas} benchmark suite for various types of fault injection (\refSec{sec:results:case_study}), and (iii) an evaluation of the accuracy of the results as we increase the count of test runs (\refSec{sec:results:runs}).

\subsection{Experimental Setup}
\label{sec:experimental_setup}
The target platform is a Linux-4.9.0, glibc-2.23  based system with an Intel\textregistered \ \mbox{Core\texttrademark \ i5-6440HQ} quad-core CPU and 8~GB RAM.
We evaluated {\nameShort} using the NAS NPB-2.3\cite{nas} benchmarks that we compiled with gcc-4.8.2\cite{gcc} using \code{-O3 -mtune=native} flags.
We used {\nameShort} 0.9.1 from \url{https://github.com/vporpo/zofi}.

\subsection{{\nameShort} Performance Overhead}
\label{sec:results:performance}

\begin{figure}[h]
\centering
\includegraphics[width=3.5in]{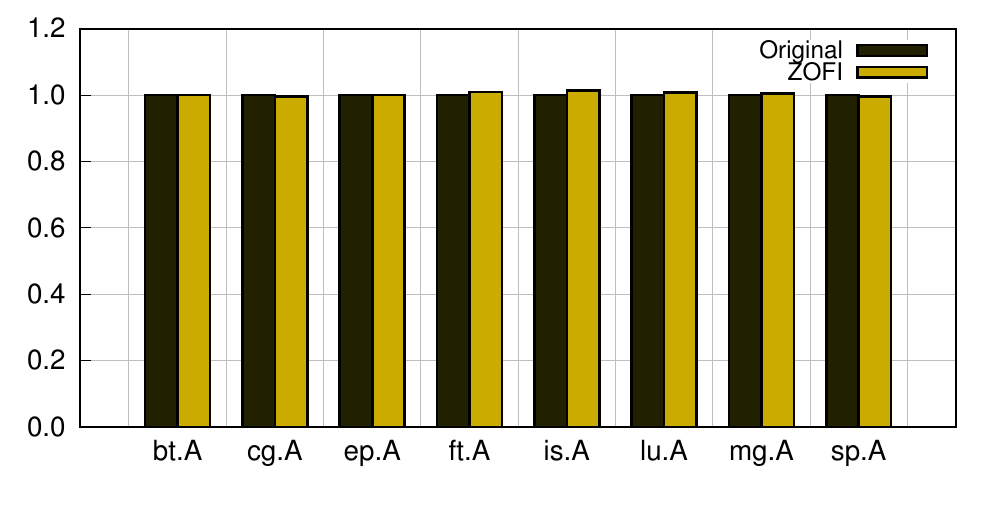}
\caption{Execution time normalized to the original (native) execution. The {\nameShort} tool introduces no performance overhead.}
\label{fig:performance_vs_raw}
\end{figure}

As already mentioned in previous sections, {\nameShort} introduces no performance overhead, because the binary is completely unmodified, and executes on the physical hardware at native speed while being \code{ptrace}d.
In order to further back our claim, we measured the execution time of the the NAS benchmarks, class A: (i) by calling them directly from the terminal, and (ii) through the {\nameShort} tool with the fault injection disabled.
The normalized results are shown in \refFig{fig:performance_vs_raw}.

It is fairly clear from the figure that there is practically no performance difference between the two measurements.
{\nameShort} does indeed execute the binaries at native speed.

We also performed a second set of experiments, where we compared the execution time of the original code versus the {\nameShort} run with an injection that lead to a masked fault.
The time taken by the {\nameShort} run, including the injection was practically identical to the original run.

\subsection{Case Study: NAS Benchmarks}
\label{sec:results:case_study}

We compiled the full NPB2.3\cite{nas} suite with the class W problem size, and then used {\nameShort} to run it, using the configuration listed in \refTab{tab:configuration}.
Each binary was executed 1000 times (\code{-test-runs 1000}), and we used all 4 cores of the target processor (\code{-j 4}).
We also used the custom diff shown in \refLst{lst:diffcmd} which filters out the output, removing those strings that vary on each run (like the throughput Mop/s, the exact execution time and the compilation date).
{\nameShort} gives us the option to choose the register types that the faults will be injected to.
For these measurements we tested five of them, as listed in \refTab{tab:configuration}, ranging from explicitly accessed  written registers (``we''), to explicitly or implicitly read or written, and with faults to the program counter enabled (``rweico'').

\begin{lstlisting}[float=*,caption={-diff-cmd for the experiment of \refSec{sec:results:case_study}.}, label={lst:diffcmd}, language=sh]
-diff-cmd /usr/bin/diff <(/bin/grep -v ``\([Tt]ime\)\|\(Mop/s total\)\|\(Compile date\)'' %ORIG_STDOUT) <(/bin/grep -v ``\([Tt]ime\)\|\(Mop/s total\)\|\(Compile date\)'' %TEST_STDOUT)
\end{lstlisting}

\begin{table}[h]
\begin{center}
  {
    \caption{{\nameShort} Configuration.}
    \begin{tabular}[t]{l|l}
      \toprule Flag & Value\\
      \midrule
\rowcolor{evencolor}      -j & 4 \\
\rowcolor{oddcolor} -test-runs & 1000 \\
\rowcolor{evencolor}-diff-cmd & See \refLst{lst:diffcmd} \\
\rowcolor{oddcolor}-inject-to & ``we''(\refFig{fig:fault_coverage_we}), ``rwe''(\refFig{fig:fault_coverage_rwe}), \\
\rowcolor{oddcolor}           & ``rwei'' (\refFig{fig:fault_coverage_rwei}), ``rweic'' (\refFig{fig:fault_coverage_rweic}), \\
\rowcolor{oddcolor}           & and ``rweico'' (\refFig{fig:fault_coverage_rweico})\\
      \midrule
    \end{tabular}

    \label{tab:configuration}
  }
\end{center}
\end{table}

\begin{figure*}[t]
\subfigure[Fault coverage with faults into explicit output registers only (``we'').\label{fig:fault_coverage_we}]{\includegraphics[width=2.25in]{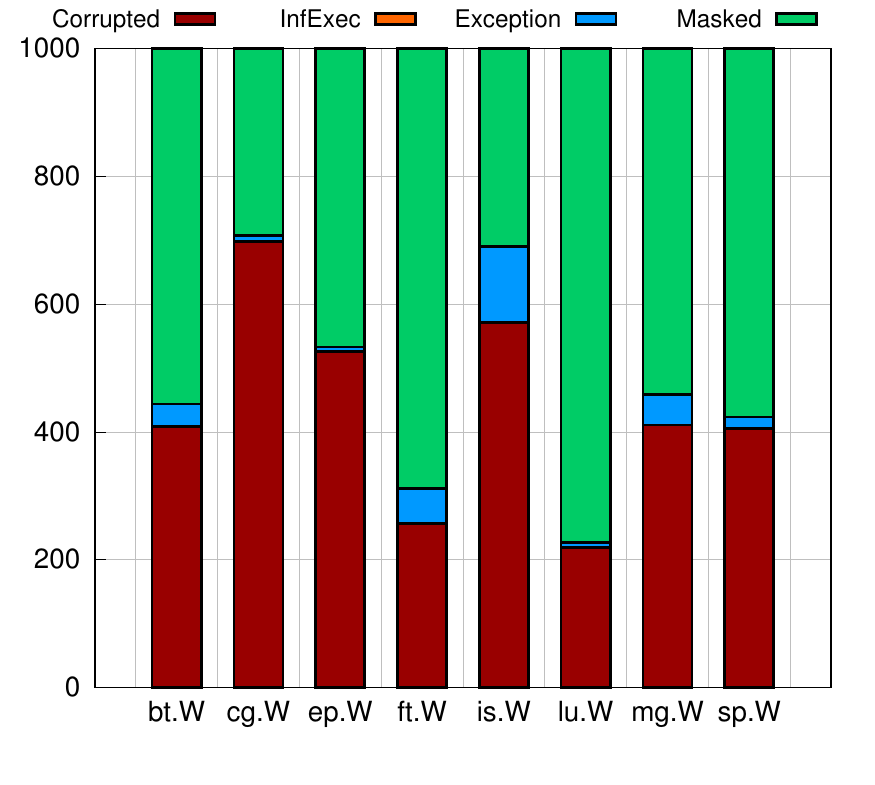}}\hspace{0.05in}
\subfigure[Fault coverage with faults into explicit input and output registers (``rwe'').\label{fig:fault_coverage_rwe}]{\includegraphics[width=2.25in]{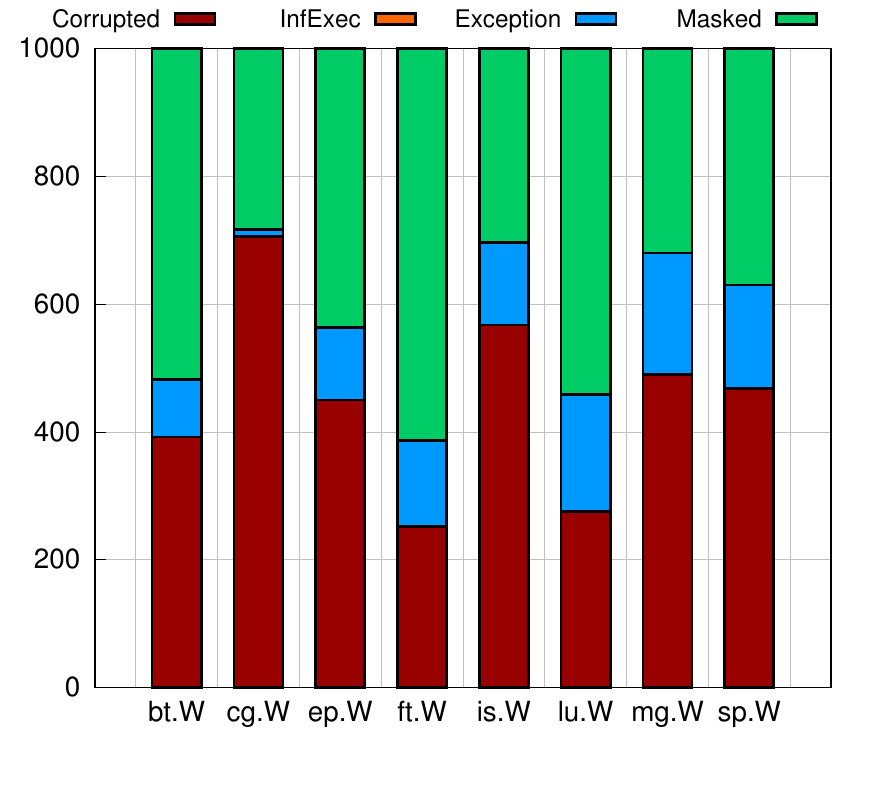}}\hspace{0.05in}
\subfigure[Fault coverage with faults into all input and output registers, including the implicitly accessed ones (``rwei'').\label{fig:fault_coverage_rwei}]{\includegraphics[width=2.25in]{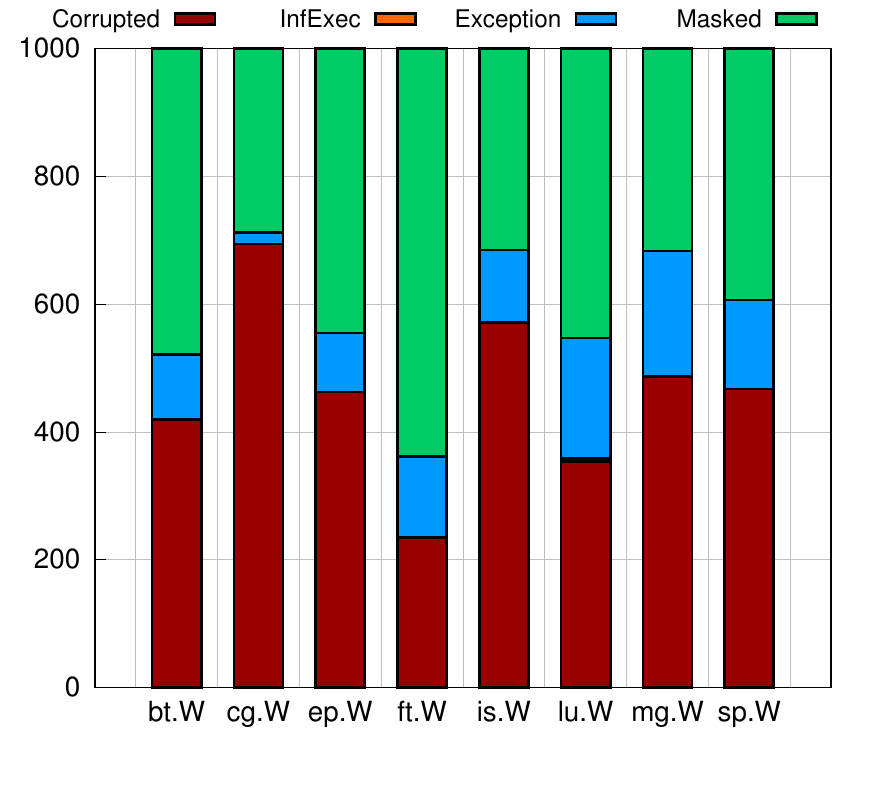}}\\

\subfigure[Fault coverage with faults into all input and output registers, including the instruction pointer for control-flow changing instructions.(``rweic'').\label{fig:fault_coverage_rweic}]{\includegraphics[width=2.25in]{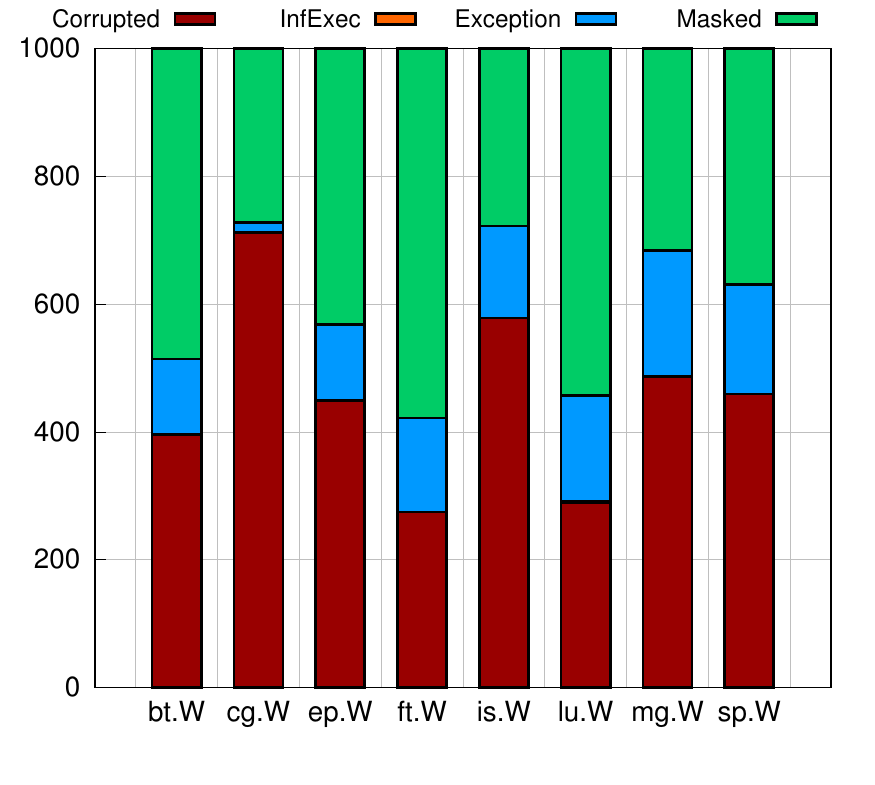}}\hspace{0.5in}
\subfigure[Fault coverage with faults into all input and output registers, including the instruction pointer for all instructions.(``rweico'').\label{fig:fault_coverage_rweico}]{\includegraphics[width=2.25in]{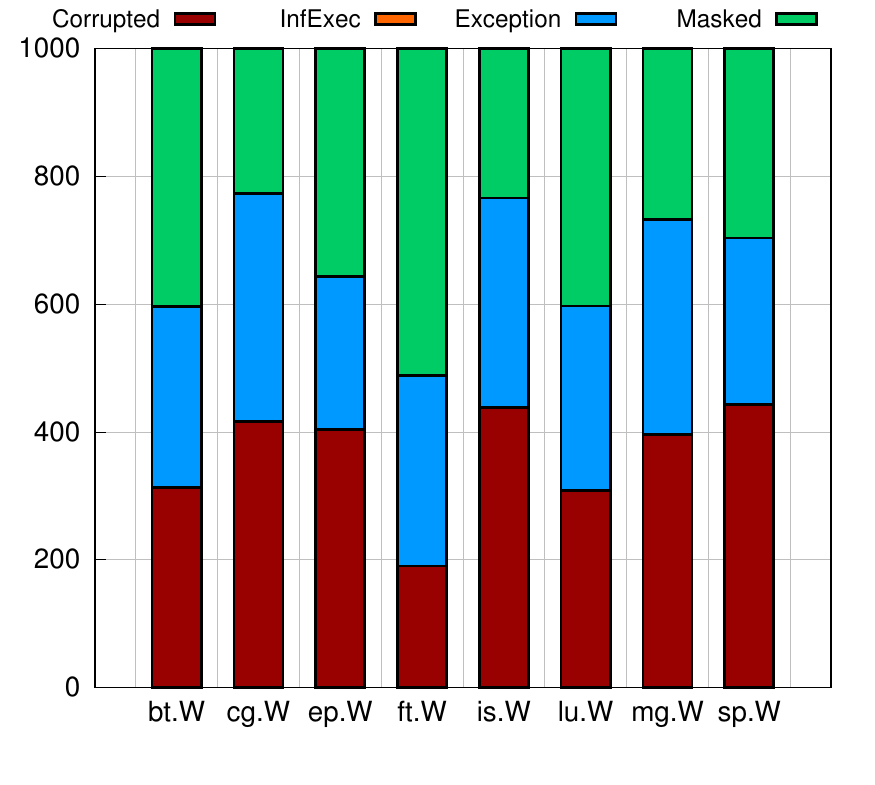}}
\caption{Fault coverage of NPB2.3, for various types of injected registers. The configuration for these tests is shown in \refTab{tab:configuration}}
\end{figure*}

For the first set of results\footnote{When checking the output of the benchmarks, we observed that is.W does not to print the ``Verification Successful'', like the rest of the benchmarks, even in the original run. Therefore we are not confident about the results for this specific benchmark.}, we injected fautls only into the explicitly accessed written registers (``we''), as shown in \refFig{fig:fault_coverage_we}.
For the second run of the benchmarks, we injected faults into both explicitly read and written registers (``rwe''), and we report the results in \refFig{fig:fault_coverage_rwe}.
Similarly, we enabled fault injection into implicit registers too (``rwei''), and report the results in \refFig{fig:fault_coverage_rwei}.
Next, we allowed control-flow instructions to cause faults to the instruction pointer (\refFig{fig:fault_coverage_rweic}), and finally, we allowed all instructions to cause faults to the instruction pointer (\refFig{fig:fault_coverage_rweico}).

Analyzing the fault coverage results is beyond the scope of this paper.
However, we will mention some very obvious points.
First of all, enabling faults in the input registers (\refFig{fig:fault_coverage_rwe}), increases the number of exceptions.
This is rather intuitive, as these registers are commonly used memory address pointers.
A bit-flip in a register that holds a memory address has a high probability of triggering a \code{SIGSEGV} signal because of the attempt to access illegal memory.
The second point is that errors in implicitly accessed registers (\refFig{fig:fault_coverage_rwei}) does not seem to have a large effect.
Given that we randomly modify a bit in the flag register, the chances of it being the one that is being used in the subsequent instruction seems rather low.
The final point is that errors in the program counter (\refFig{fig:fault_coverage_rweico}) cause a dramatic increase in the number of exceptions.
This is expected, as faults in the program counter will likely cause a jump to an illegal address, which will trigger an exception.

\subsection{Accuracy of Results as we Increase the Number of Test Runs}
\label{sec:results:runs}
It is a well known fact that the higher number of the test runs in a Monte Carlo approach, the better the accuracy of the results.
Most studies in the software-based error-detection literature use a few hundreds (e.g., \cite{khudia2013low,drift}) to a few thousands (e.g., \cite{ghosh2012runtime}) of runs for the fault coverage evaluation.
The number of runs is usually selected in an arbitrary manner.
In this section we are looking into the error that one can expect across independent runs of {\nameShort}, for a couple of benchmarks of the NPB2.3 suite.
We selected cg.W, ft.W and mg.W because these run the fastest among all benchmarks in the suite.
To this end, we varied the number of program runs from 50 to 6400, running each configuration 10 times, and calculating the arithmetic mean and standard deviation of the results across these 10 repetitions.
This is a total of 382500 runs, which took about a day to run on our i5-6440HQ quad-core system.
The configurations that we tested are summarized in \refTab{tab:configuration:runs}.

\begin{table}[h]
\begin{center}
  {
    \caption{{\nameShort} Configuration for results of \refSec{sec:results:runs}.}
    \begin{tabular}[t]{l|l}
      \toprule Flag & Value\\
      \midrule
\rowcolor{evencolor}      -j & 4 \\
\rowcolor{oddcolor}-test-runs & 50, 100, 200, 400, 800, 1600, 3200, 6400 \\
\rowcolor{evencolor}-inject-to & ``rwe'' \\
\rowcolor{oddcolor}-diff-cmd & See \refLst{lst:diffcmd} \\
      \midrule
    \end{tabular}

    \label{tab:configuration:runs}
  }
\end{center}
\end{table}

\begin{figure*}[t]
\centering
\subfigure[cg.W\label{fig:fault_coverage_multiple_runs_cg}]{\includegraphics[width=2.3in]{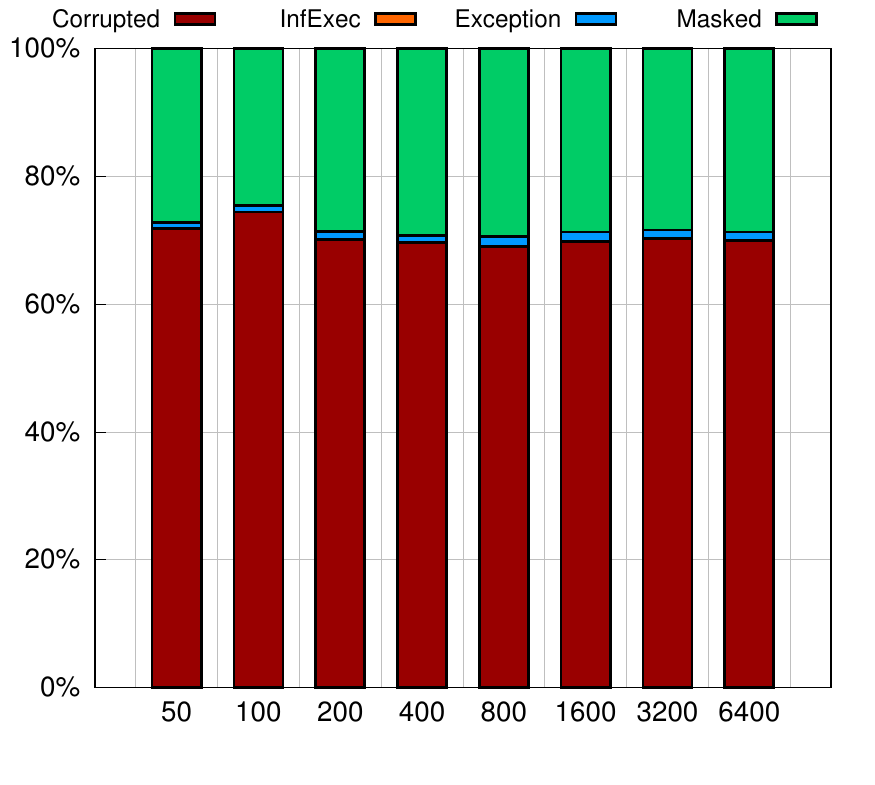}} 
\subfigure[ft.W\label{fig:fault_coverage_multiple_runs_ft}]{\includegraphics[width=2.3in]{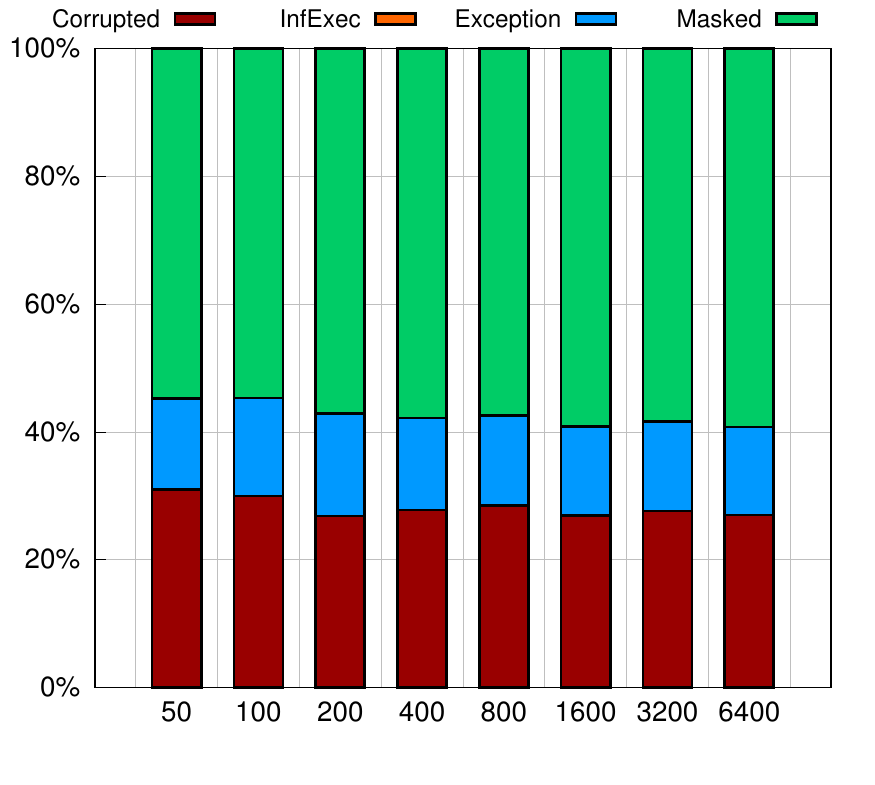}} 
\subfigure[mg.W\label{fig:fault_coverage_multiple_runs_mg}]{\includegraphics[width=2.3in]{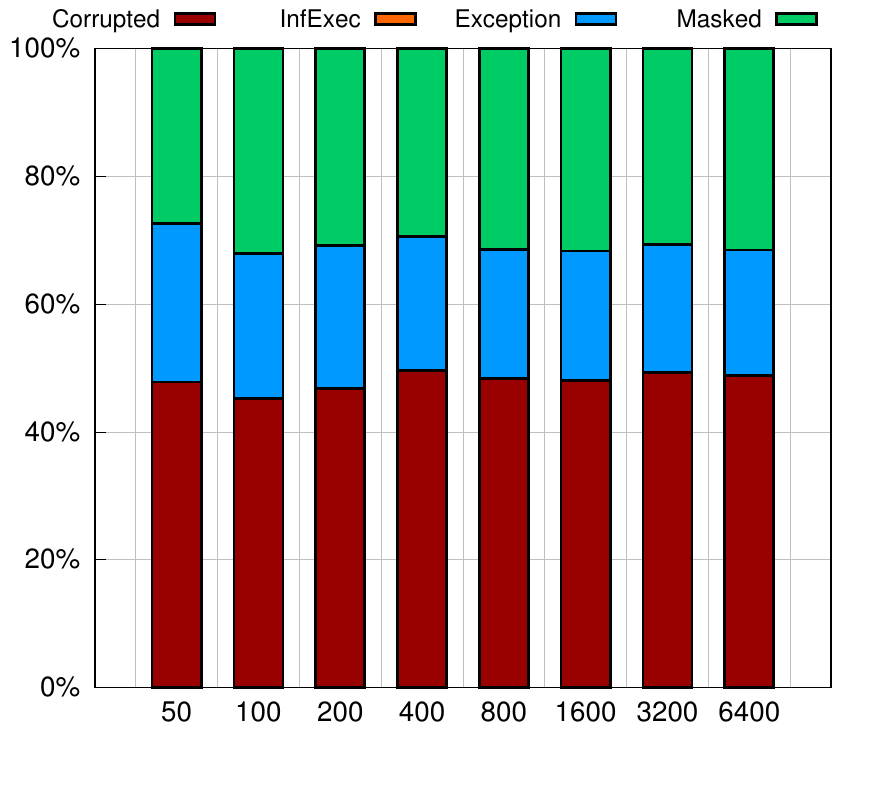}}
\caption{Mean fault coverage (\%) across 10 repetitions of {\nameShort}, as we vary the number of runs from 50 to 6400, injecting into explicitly written registers (``rwe'').}
\label{fig:fault_coverage_multiple_runs_mean}
\end{figure*}

\begin{figure*}[t]
\subfigure[cg.W\label{fig:fault_coverage_multiple_runs_err_cg}]{\includegraphics[width=2.3in]{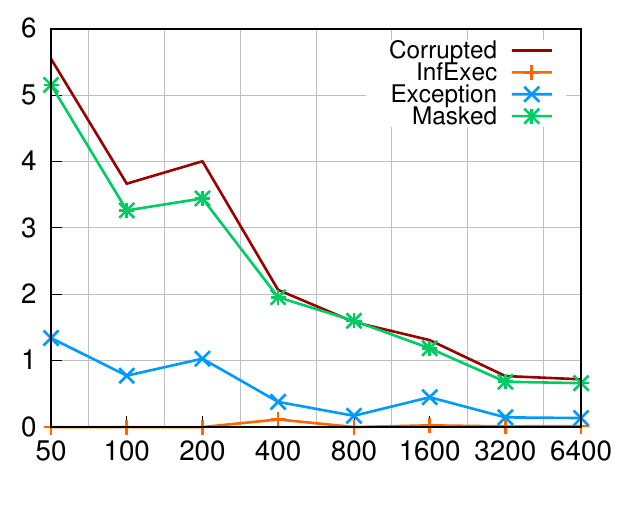}}
\subfigure[ft.W\label{fig:fault_coverage_multiple_runs_err_ft}]{\includegraphics[width=2.3in]{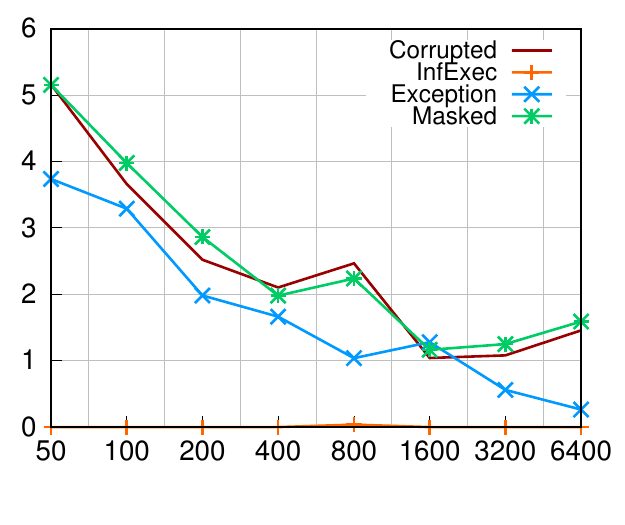}}
\subfigure[mg.W\label{fig:fault_coverage_multiple_runs_err_mg}]{\includegraphics[width=2.3in]{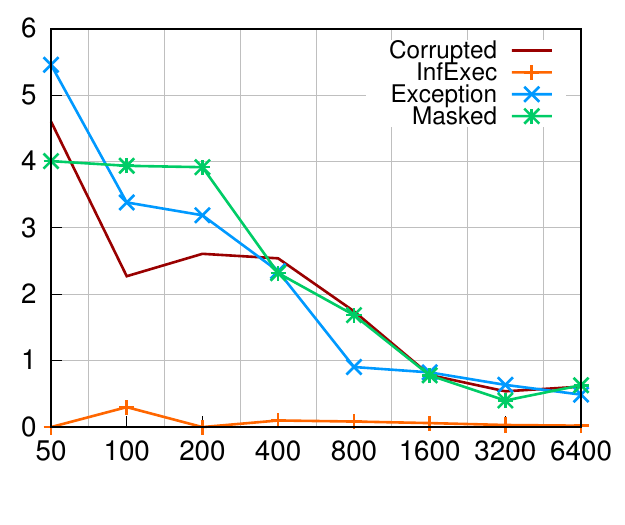}}
\caption{Standard deviation of the fault coverage results across 10 repetitions of {\nameShort}, as we vary the number of runs from 50 to 6400. The mean values of these experiments are shown in \refFig{fig:fault_coverage_multiple_runs_mean}.}
\label{fig:fault_coverage_multiple_runs_err}
\end{figure*}

\refFigss{fig:fault_coverage_multiple_runs_cg}{fig:fault_coverage_multiple_runs_ft}{fig:fault_coverage_multiple_runs_mg} show the arithmetic mean of the fault coverage results for cg.W, ft.W and mg.W, as we increase the test runs from 50 to 6400.
In other words, each run of the {\nameShort} tool reports the percentages of each outcome.
These results will vary across runs, so we collect this data across all 10 runs and we report the arithmetic mean in \refFigss{fig:fault_coverage_multiple_runs_cg}{fig:fault_coverage_multiple_runs_ft}{fig:fault_coverage_multiple_runs_mg} and the standard deviation in \refFigss{fig:fault_coverage_multiple_runs_err_cg}{fig:fault_coverage_multiple_runs_err_ft}{fig:fault_coverage_multiple_runs_err_mg}.
Each line in the standard-deviation plots represents the outcome (i.e., Corruption, Infinite Execution, Exception and Masked).

What is particularly interesting, is that the arithmetic mean across 10 runs of {\nameShort} seems to be quite accurate even for just 50 test runs of the workload. 
However, the error across these 10 runs of the tool is quite high, meaning that we could not trust the fault coverage reported by a single 50-test run of the tool.

As expected, the standard deviation decreases as we increase the number of runs. This is clearly shown in \refFigss{fig:fault_coverage_multiple_runs_err_cg}{fig:fault_coverage_multiple_runs_err_ft}{fig:fault_coverage_multiple_runs_err_mg}.
There is, however, some noise in these results, particularly towards smaller values of the number of test-runs, which would probably improve if we ran the tool more than 10 times.
We can also observe that the decrease of the error is rather small compared to the increase of the number of runs.
Please note that the horizontal axis of these plots is in logarithmic scale, and even so the error deviation decreases at a diminishing rate.

\section{Conclusion}
This work presented {\nameShort}, a timing-based Zero-Overhead Fault Injection tool for very fast fault coverage evaluation at native speed.
Unlike the frameworks used in the majority of software-based error detection studies, {\nameShort} does not instrument the binary nor does it run it on a simulator.
Instead, it runs the unmodified workload directly on native hardware, with no performance overhead whatsoever.
It is a timing-based tool, meaning that it operates at time-points, not at cycles.
It collects fault-coverage statistics across multiple test runs and reports their summary upon completion.
{\nameShort} is distributed as free software and is available at \url{https://github.com/vporpo/zofi}.
\vspace{0.7in}


\end{document}